\documentclass[english,]{IEEEtran}

\usepackage{lmodern}
\usepackage{amssymb,amsmath}
\usepackage{ifxetex,ifluatex}
\usepackage{fixltx2e} 
\ifnum 0\ifxetex 1\fi\ifluatex 1\fi=0 
  \usepackage[T1]{fontenc}
  \usepackage[utf8]{inputenc}
\else 
  \ifxetex
    \usepackage{mathspec}
  \else
    \usepackage{fontspec}
  \fi
  \defaultfontfeatures{Ligatures=TeX,Scale=MatchLowercase}
\fi
\IfFileExists{upquote.sty}{\usepackage{upquote}}{}
\IfFileExists{microtype.sty}{%
\usepackage{microtype}
\UseMicrotypeSet[protrusion]{basicmath} 
}{}
\usepackage[unicode=true]{hyperref}
\hypersetup{
            pdftitle={SkiffOS: Minimal Cross-compiled Linux for Embedded Containers},
            pdfkeywords={Container, Embedded Linux, Robotics, Single
Board Computer},
            pdfborder={0 0 0},
            breaklinks=true}
\ifnum 0\ifxetex 1\fi\ifluatex 1\fi=0 
  \usepackage[shorthands=off,main=english]{babel}
\else
  \usepackage{polyglossia}
  \setmainlanguage[]{english}
\fi
\usepackage{csquotes}
\usepackage[backend=biber,style=ieee,natbib=true]{biblatex} 

\usepackage{color}
\usepackage{fancyvrb}

\DefineVerbatimEnvironment{Highlighting}{Verbatim}{commandchars=\\\{\}}
\newenvironment{Shaded}{}{}

\newcommand{\AttributeTok}[1]{\textcolor[rgb]{0.49,0.56,0.16}{#1}}

\newcommand{\CharTok}[1]{\textcolor[rgb]{0.25,0.44,0.63}{#1}}

\newcommand{\FunctionTok}[1]{\textcolor[rgb]{0.02,0.16,0.49}{#1}}

\newcommand{\KeywordTok}[1]{\textcolor[rgb]{0.00,0.44,0.13}{\textbf{#1}}}

\usepackage{longtable,booktabs}
\usepackage{supertabular}

\let\endhead\relax
\IfFileExists{footnote.sty}{\usepackage{footnote}\makesavenoteenv{long table}}{}
\IfFileExists{parskip.sty}{%
\usepackage{parskip}
}{
\setlength{\parindent}{0pt}
\setlength{\parskip}{6pt plus 2pt minus 1pt}
}
\providecommand{\tightlist}{%
  \setlength{\itemsep}{0pt}\setlength{\parskip}{0pt}}
\ifx\paragraph\undefined\else
\let\oldparagraph\paragraph
\renewcommand{\paragraph}[1]{\oldparagraph{#1}\mbox{}}
\fi
\ifx\subparagraph\undefined\else
\let\oldsubparagraph\subparagraph
\renewcommand{\subparagraph}[1]{\oldsubparagraph{#1}\mbox{}}
\fi

\makeatletter
\def\fps@figure{htbp}
\makeatother

\title{SkiffOS: Minimal Cross-compiled Linux for Embedded Containers}

\author{
            \IEEEauthorblockN{Christian Stewart}
        \IEEEauthorblockA{%
             \\
            Aperture Robotics LLC. \\
            christian@aperturerobotics.com}
        }

\date{}

\begin{document}

\maketitle
\begin{abstract}
Embedded Linux processors are increasingly used for real-time computing
tasks such as robotics and Internet of Things (IoT). These applications
require robust and reproducible behavior from the host OS, commonly
achieved through immutable firmware stored in read-only memory. SkiffOS
addresses these requirements with a minimal cross-compiled GNU/Linux
system optimized for hosting containerized distributions and
applications, and a configuration layering system for the Buildroot
embedded cross-compiler tool which automatically re-targets system
configurations to any platform or device. This approach cleanly
separates the hardware support from the applications. The host system
and containers are independently upgraded and backed-up Over-the-Air
(OTA).
\end{abstract}


\hypertarget{introduction}{%
\section{Introduction}\label{introduction}}

Single Board Computer (SBC) vendors distribute "board support packages"
(BSPs) containing a mutable root filesystem with a package manager. Each
platform requires a unique combination of kernel, bootloader, and
firmware to operate, as a result BSPs typically lack portability.

Most Linux distributions use binary package managers to install and
upgrade applications and system files, managed by distribution
maintainers. Application dependencies co-exist with system dependencies,
often creating version conflicts resulting in out-of-date and
potentially vulnerable system software \autocite{foss}. Package
availability and versioning can vary across processor architectures.

Package managers operate on a mutable filesystem with imperative
management commands (i.e.~install, uninstall, upgrade) \autocite{nixos}.
Each modification to the system configuration diverges from the known
initial state, and can result in unpredictable changes in system
behavior such as broken boot, inaccessible ssh, or malfunctioning
drivers.

Declarative configuration models allow for reproducible behavior from
the OS. Embedded Linux systems such as IoT sensors, industrial
automation, TV set-top boxes, webcams, kiosks, and other products often
bundle a GNU/Linux system into read-only firmware based on declarative
configurations. This guarantees reproducible behavior, but comes with
significant disadvantages in usability: software typically cannot be
installed or upgraded without a system re-build and re-flash.

\hypertarget{skiffos-firmware-for-hosting-containers}{%
\section{SkiffOS: Firmware for Hosting
Containers}\label{skiffos-firmware-for-hosting-containers}}

Buildroot is a project providing a Makefile and KConfig-based tool for
automated cross-compilation of GNU/Linux systems, including toolchains,
kernels, drivers, userspaces, and bootloaders \autocite{buildroot}. It
includes over 2500 software packages providing support for all major
system utilities, and provides an easy to understand structure and
configuration language allowing for rapid development across a diverse
set of hardware configurations \autocite{buildrootrt}.

SkiffOS builds a minimal cross-compiled system for hosting containers
across a diverse set of compute platforms, with minimal variance in the
user experience. It emulates traditional firmware approaches with an
immutable operating system image which produces reproducible behavior,
while providing the utility of package-manager based GNU/Linux
distributions with containerized environments attached to persistent
storage.

Configuration layers bundle configuration files, Buildroot extension
packages, system files, build hooks, and installation/utility scripts
into named packages. Configurations are supplied for numerous target
systems and virtualization environments. Hardware-specific performance
optimizations are enabled at build time, producing a system image tuned
for the target computer. Makefile targets are available for formatting
and installing the system to boot media. Multiple system configurations
can be compiled in parallel with "workspaces."

The SkiffOS Git repository can be embedded in extension projects as a
sub-module and extended with out-of-tree configuration packages and
overrides. Buildroot is referenced as a sub-module within the SkiffOS
Git tree. Checksumming, package version pinning, and reproducible
offline builds are used to ensure that a given SkiffOS commit will
always produce identical output.

Existing system userspaces can be imported and used directly as
container images Package managers can then be used to install and manage
software independently from the host system. Multiple distributions and
containerized applications can be run in parallel. Workloads can be
defined as container images to enhance portability and reproducibility
\autocite{reworkflow}. Container management platforms such as Kubernetes
and Docker Swarm can be used to remotely deploy and monitor workloads
\autocite{autoveh}.

\hypertarget{implementation}{%
\section{Implementation}\label{implementation}}

SkiffOS is available under the MIT license, and references Buildroot as
a sub-module under the GPLv2 license, with a patch series providing
additional features and bug fixes. Changes are frequently submitted
upstream to the Buildroot mailing list.

As of Release 2020.11.7 the target support table is:

\begin{longtable}[]{@{}lll@{}}
\toprule
System & Config Package & Kernel \\ \addlinespace
\midrule
\endhead
Apple Macbook (Intel) & apple/macbook & 5.11.2 \\ \addlinespace
BananaPi M1 & bananapi/m1 & 5.11.2 \\ \addlinespace
BananaPi M1+/Pro & bananapi/m1plus & 5.11.2 \\ \addlinespace
BananaPi M2+ & bananapi/m2plus & 5.11.2 \\ \addlinespace
BananaPi M3 & bananapi/m3 & 5.11.2 \\ \addlinespace
Docker Container & virt/docker & N/A \\ \addlinespace
Intel x86/64 & intel/x64 & 5.11.2 \\ \addlinespace
Msft Windows (WSL) & virt/wsl & N/A \\ \addlinespace
NVIDIA Jetson Nano & jetson/nano & 4.9.140 \\ \addlinespace
NVIDIA Jetson TX2 & jetson/tx2 & 4.9.140 \\ \addlinespace
Odroid C2 & odroid/c2 & tb-5.9.16 \\ \addlinespace
Odroid C4 & odroid/c4 & tb-5.9.16 \\ \addlinespace
Odroid HC1/2, XU3/4 & odroid/xu & tb-5.9.16 \\ \addlinespace
Odroid U & odroid/u & tb-5.9.16 \\ \addlinespace
OrangePi Lite & orangepi/lite & 5.11.2 \\ \addlinespace
OrangePi Zero & orangepi/zero & 5.11.2 \\ \addlinespace
PcDuino 3 & pcduino/3 & 5.11.2 \\ \addlinespace
PcEngines APU2 & pcengines/apu2 & 5.11.2 \\ \addlinespace
Pi 0 & pi/0 & 5.4.72 \\ \addlinespace
Pi 1 & pi/1 & 5.4.72 \\ \addlinespace
Pi 3 (and 1/2) & pi/3 & 5.4.72 \\ \addlinespace
Pi 4 & pi/4 & 5.4.72 \\ \addlinespace
Pine64 H64 & pine64/h64 & 5.8.0 \\ \addlinespace
PinePhone & pine64/phone & megi-5.9.11 \\ \addlinespace
Pinebook Pro & pine64/book & 5.11.2 \\ \addlinespace
Qemu (VM) & virt/qemu & 5.11.2 \\ \addlinespace
Virtualbox (VM) & virt/qemu & 5.11.2 \\ \addlinespace
Rockpro64 & pine64/rockpro64 & 5.9.0 \\ \addlinespace
\bottomrule
\end{longtable}

Legal information and licenses for all dependencies can be bundled
together with the \texttt{make\ legal-info} command. Some board support
packages include proprietary binary blobs (typically firmware) and are
denoted as "Proprietary" in the produced licensing information bundle.

\hypertarget{configuration-layers}{%
\subsection{Configuration Layers}\label{configuration-layers}}

The SkiffOS base configuration, builds a minimal host operating system
with hardware support, OpenSSH server, and system management tools.
Enabling additional configuration layers adds use-case specific
functionality. Configurations are organized into logical units called
"layers" with the following structure:

\begin{itemize}
\tightlist
\item
  \texttt{cflags}: additional target compiler flags
\item
  \texttt{buildroot}: buildroot configuration fragments
\item
  \texttt{buildroot\_ext}: buildroot extensions
\item
  \texttt{buildroot\_patches}: buildroot package patches
\item
  \texttt{extensions}: utility commands
\item
  \texttt{hooks}: scripts hooking pre/post build steps
\item
  \texttt{kernel}: kernel configuration fragments
\item
  \texttt{kernel\_patches}: kernel .patch files
\item
  \texttt{root\_overlay}: root overlay files
\item
  \texttt{metadata}: metadata files

  \begin{itemize}
  \tightlist
  \item
    \texttt{commands}: targets in "extensions" makefile
  \item
    \texttt{dependencies}: comma-separated layers
  \item
    \texttt{description}: single-line description
  \item
    \texttt{unlisted}: if exists, hidden from "help"
  \end{itemize}
\item
  \texttt{resources}: support files
\item
  \texttt{scripts}: used by extensions and/or hooks
\item
  \texttt{uboot}: u-boot configuration fragments
\item
  \texttt{uboot\_patches}: u-boot .patch files
\end{itemize}

Configuration layers can override options set by previous layers, making
it simple to re-target configurations to various compute platforms by
merging with the desired hardware configuration layer. The set of
desired configuration layers is defined as an ordered comma-separated
list, for example:

\texttt{SKIFF\_CONFIG=pi/4,core/gentoo}

SkiffOS can be extended \autocite{skiffext} by adding it as a sub-module
of a project repo. Project configurations can then be specified in
additional configuration layers.

\hypertarget{persistent-data-partition}{%
\subsection{Persistent data partition}\label{persistent-data-partition}}

The persist partition contains a \texttt{skiff} directory with:

\begin{itemize}
\tightlist
\item
  \texttt{connections}: network-manager connections
\item
  \texttt{core}: skiff-core configuration and state
\item
  \texttt{docker}: docker state and storage
\item
  \texttt{etc}: configuration tree overlay
\item
  \texttt{hostname}: system hostname
\item
  \texttt{journal}: systemd-journald system logs
\item
  \texttt{keys}: public keys for access to "root" user
\item
  \texttt{ssh}: ssh server keys and persistent configuration
\end{itemize}

System startup scripts mount the persist partition and create the file
structure at boot time. Most configurations will create a memory
swapfile to avoid failure due to an out-of-memory condition. OpenSSH is
configured for public-key access with the persist "keys" directory or in
the OS image.

The persist partition is automatically resized to fit the remainder of
the available storage space on first boot. Bootloaders such as u-boot
are sometimes copied to the beginning of the storage media. Some systems
use more complex partition layouts mandated by the hardware vendor.
Skiff and Buildroot's flexible configuration language supports this
variance between target systems.

\hypertarget{over-the-air-ota-upgrades}{%
\subsection{Over-the-air (OTA)
upgrades}\label{over-the-air-ota-upgrades}}

The SkiffOS system typically consists of fewer than five files,
including the root filesystem squashfs/initramfs, kernel image, and
kernel modules squashfs. The kernel and immutable boot system can be
atomically upgraded by replacing these files at run-time. The
\texttt{push\_image.sh} script is provided, using \texttt{rsync} and
\texttt{ssh} to upload the updated files to a running system. In some
cases, the firmware and/or bootloader is also upgraded by the script.

\texttt{./scripts/push\_image.sh\ root@my-device-ip}

\hypertarget{kconfig-configuration-fragments}{%
\subsection{Kconfig configuration
fragments}\label{kconfig-configuration-fragments}}

Buildroot organizes software components into "packages." Packages can be
enabled and configured using the Kconfig language. The available options
can be explored using the \texttt{make\ menuconfig} or
\texttt{make\ xconfig} configuration menus. Configuration options are
specified in "fragments" that are merged together into a single
".config" file and provided to the Buildroot build system. The Buildroot
build system manages merging together kconfig and u-boot configuration
fragments and applying patches.

Example buildroot configuration fragment:

\begin{verbatim}
BR2_LINUX_KERNEL_DEFCONFIG="versatile"
BR2_LINUX_KERNEL_CUSTOM_VERSION_VALUE="5.11.2"
\end{verbatim}

Buildroot and the Linux kernel use the Kconfig configuration language.
The available options can be explored with \texttt{make\ br/menuconfig}
and \texttt{make\ br/kernel-menuconfig} configuration menus.
Configuration fragments are merged together in \texttt{SKIFF\_CONFIG}
order followed by lexicographic filename order.

This example kernel configuration fragment enables the ext3/ext4
filesystem:

\begin{verbatim}
CONFIG_EXT3_FS=m
CONFIG_EXT3_FS_SECURITY=y
CONFIG_EXT3_FS_XATTR=y
CONFIG_EXT3_POSIX_ACL=y
CONFIG_EXT4_FS=y
CONFIG_EXT4_FS_SECURITY=y
CONFIG_EXT4_POSIX_ACL=y
\end{verbatim}

The \texttt{m} option denotes a feature built as a kernel module instead
of "built-in."

\hypertarget{buildroot-extensions}{%
\subsection{Buildroot extensions}\label{buildroot-extensions}}

Configuration layers can extend Buildroot with packages in the
\texttt{buildroot\_ext} directory. The \texttt{buildroot\_patches}
directory contains patches for Buildroot packages, i.e.~fixes for
platform errata. For convenience, kernel and uboot patches can also be
specified in the \texttt{kernel\_patches} and \texttt{uboot\_patches}
directories.

\hypertarget{skiffos-extensions}{%
\subsection{SkiffOS extensions}\label{skiffos-extensions}}

The \texttt{extensions} configuration layer directory contains a
Makefile which implements custom commands made available to the user in
the help screen:

\begin{verbatim}
cmd/pi/common/format: Format a SD card.
cmd/pi/common/install: Install to a SD card.
\end{verbatim}

Commands are prefixed by their layer name and can be executed as
Makefile targets, for example, \texttt{make\ cmd/pi/common/install}, and
are declared in the \texttt{metadata/commands} text file:

\begin{verbatim}
format Format a SD card and install bootloader.
install Installs to a formatted SD card.
\end{verbatim}

\hypertarget{root-filesystem-overlay}{%
\subsection{Root filesystem overlay}\label{root-filesystem-overlay}}

The \texttt{root\_overlay} trees are copied to the target filesystem
image at the end of the build in the order that the layers were
specified. This is used to add additional configuration files or scripts
to the root SkiffOS system image. For example, the \texttt{pi/common}
layer adds configuration under the \texttt{etc} directory, and firmware
configuration under the \texttt{usr} directory.

\hypertarget{temporary-local-overrides}{%
\subsection{Temporary local overrides}\label{temporary-local-overrides}}

Configuration overrides can be specified in the \texttt{overrides}
directory tree, which contains additional configuration layers
implicitly added to the build. For example, temporary local buildroot
overrides for all workspaces can be declared in
\texttt{overrides/buildroot}, and kernel configuration fragments
affecting the \texttt{pi4} workspace alone can be declared in
\texttt{overrides/workspaces/pi4/kernel}.

\hypertarget{skiff-core-containerized-environments}{%
\section{Skiff Core: Containerized
Environments}\label{skiff-core-containerized-environments}}

Skiff Core includes the Docker containerization runtime and a Go program
for automating the creation and initial setup of containerized
environments. It can be enabled with the \texttt{skiff/core}
configuration layer. User sessions are routed to the container assigned
to their account. Multiple OS distributions can be installed
simultaneously on a single machine. Existing userspaces including
vendor-provided software images can be imported as container images.

The \texttt{skiff-core} binary is configured as the user shell, and
intercepts incoming SSH sessions to redirect them to the corresponding
container. The container setup process is displayed if the container(s)
are not yet ready. The usual userspace init daemon, typically
\texttt{systemd}, is run as PID 1, and standard approaches for defining
\texttt{systemd} services and the \texttt{systemctl} CLI tool function
identically to when running without containerization.

The "core" system can be updated or rolled-back independently from the
"root" operating system. Mountpoints are used to mount user home
directories and other temporary paths so that "docker export" and
"docker save" include system files only. Containers are portable between
machines of similar architecture without target-specific configuration.

The container system is configured with a YAML file:

\begin{Shaded}
\begin{Highlighting}[]
\FunctionTok{containers}\KeywordTok{:}
\AttributeTok{  }\FunctionTok{core}\KeywordTok{:}
\AttributeTok{    }\FunctionTok{image}\KeywordTok{:}\AttributeTok{ skiffos/skiff{-}core{-}gentoo:latest}
\AttributeTok{    }\FunctionTok{mounts}\KeywordTok{:}
\AttributeTok{      }\KeywordTok{{-}}\AttributeTok{ /dev:/dev}
\AttributeTok{      }\KeywordTok{{-}}\AttributeTok{ /etc/resolv.conf:/etc/resolv.conf:ro}
\AttributeTok{      }\KeywordTok{{-}}\AttributeTok{ /mnt/persist/data:/home}
\AttributeTok{    }\KeywordTok{[}\AttributeTok{...}\KeywordTok{]}
\FunctionTok{users}\KeywordTok{:}
\AttributeTok{  }\FunctionTok{core}\KeywordTok{:}
\AttributeTok{    }\FunctionTok{container}\KeywordTok{:}\AttributeTok{ core}
\AttributeTok{    }\FunctionTok{containerUser}\KeywordTok{:}\AttributeTok{ core}
\AttributeTok{    }\FunctionTok{auth}\KeywordTok{:}\AttributeTok{ }\KeywordTok{\{}\FunctionTok{copyRootKeys}\KeywordTok{:}\AttributeTok{ }\CharTok{true}\KeywordTok{\}}
\FunctionTok{images}\KeywordTok{:}
\AttributeTok{  skiffos/skiff{-}core{-}gentoo}\FunctionTok{:latest}\KeywordTok{:}
\AttributeTok{    }\FunctionTok{pull}\KeywordTok{:}
\AttributeTok{      }\FunctionTok{policy}\KeywordTok{:}\AttributeTok{ ifnotexists}
\AttributeTok{      }\FunctionTok{registry}\KeywordTok{:}\AttributeTok{ quay.io}
\AttributeTok{    }\FunctionTok{build}\KeywordTok{:}
\AttributeTok{      }\FunctionTok{source}\KeywordTok{:}\AttributeTok{ /opt/skiff/coreenv/base}
\end{Highlighting}
\end{Shaded}

Several OS distribution configurations are available as configuration
layers. If a pre-built image is unavailable, or the \texttt{pull}
section of \texttt{skiff-core.yaml} is empty, the system will instead
build the included \texttt{Dockerfile}.

\begin{longtable}[]{@{}lll@{}}
\toprule
OS Distribution & Config Package & Website \\ \addlinespace
\midrule
\endhead
Gentoo &
\href{https://github.com/skiffos/skiffos/tree/2020.11.7/configs/core/gentoo}{core/gentoo}
& \href{https://gentoo.org}{gentoo.org} \\ \addlinespace
Manjaro &
\href{https://github.com/skiffos/skiffos/tree/2020.11.7/configs/core/manjaro}{core/manjaro}
& \href{https://manjaro.org}{manjaro.org} \\ \addlinespace
NixOS &
\href{https://github.com/skiffos/skiffos/tree/2020.11.7/configs/core/nixos/}{core/nixos}
& \href{https://nixos.org}{nixos.org} \\ \addlinespace
Ubuntu &
\href{https://github.com/skiffos/skiffos/tree/2020.11.7/configs/skiff/core/}{skiff/core}
& \href{https://ubuntu.org}{ubuntu.org} \\ \addlinespace
\bottomrule
\end{longtable}

Additional images optimized for specific use cases are available:

\begin{longtable}[]{@{}ll@{}}
\toprule
Image & Description \\ \addlinespace
\midrule
\endhead
gentoo-lto-exwm & LTO, Apps, Emacs X11 Workflow \\ \addlinespace
gentoo-kde & KDE Desktop w/ apps \\ \addlinespace
gentoo-lto & O3, Graphite, Link-time optimize \\ \addlinespace
nasa-cfs & Flight systems framework \\ \addlinespace
nasa-fprime & Flight systems framework \\ \addlinespace
pinephone-neon & KDE Neon for PinePhone \\ \addlinespace
-manjaro-kde & Manjaro KDE for PinePhone \\ \addlinespace
\bottomrule
\end{longtable}

The NASA Fprime\autocite{fprime} and NASA cFS\autocite{nasacfs} images
are currently based on Ubuntu. To pull, for example, the
\texttt{gentoo-lto} image, which includes the Gentoo core image with
\texttt{gentooLTO} optimizations, the Docker CLI command is:

\texttt{docker\ pull\ skiffos/skiff-core-gentoo-lto:latest}

When using "Skiff Core," the Docker container engine is used, which
leverages Linux namespaces and not virtualization or emulation,
implementing "OS-level virtualization as opposed to hardware
virtualization" \autocite{virtperf}. The PID namespace is used to allow
running the usual system initialization and management daemons as the
privileged PID 1 within the container.

The results of \autocite{iotcontainer} show "an almost negligible impact
of the {[}Docker container{]} layer in terms of performance, if compared
to native execution." While throughput is not significantly affected,
the results of \autocite{autoveh} show that namespacing can cause a
measurable signal processing delay. To mitigate potential impacts of
processing latency, most container isolation is disabled.

\hypertarget{comparison-with-existing-approaches}{%
\section{Comparison with Existing
Approaches}\label{comparison-with-existing-approaches}}

This section discusses several of the current approaches in use today,
and their disadvantages when applied to embedded Linux and/or robotics
development:

\hypertarget{binary-package-distributions}{%
\subsubsection{Binary Package
Distributions}\label{binary-package-distributions}}

Traditional board-support packages use binary package distribution
systems such as Debian's Advanced Package Tool (APT). Several key
disadvantages of this approach are lack of hardware-specific
optimizations, reliance on third-party infrastructure for builds and
maintenance, inability to reproduce a system from source code, and
difficulties with portability. The rolling nature of package upgrades
introduces unpredictable behavior when upgrading a system, particularly
when a long time has passed since the previous upgrade \autocite{foss}.

Current widely-used package management tools install system files,
firmware, and user applications together into a single mutable "root"
filesystem. Imperative install, remove, and upgrade commands instruct
the package manager to perform operations on the system. Interrupted
package manager operations might leave some files in a partially written
state. If the affected (now corrupted) files are critical to system boot
and/or reachability, the only recourse to fix the system may be to
remove the boot media from the machine, connect it to a different
device, and fix the issue manually.

SkiffOS addresses the concern of always having reproducible boot-up and
reachability behavior in embedded systems by booting to an immutable
root filesystem image, mounting persistent storage, and running user
applications and operating systems inside lightweight Linux containers.
This mitigates the risks of mid-upgrade power brownout: the immutable
portion of the system is always reachable, allowing the user can connect
to the container and fix the problem without physical intervention.

Optimization of compute performance and reduction of energy usage are
important considerations for resource-constrained embedded devices (the
"power budget"). For most robotics applications, the "power budget" is a
significant factor in determining maximum range/endurance. Compilation
of the operating system and applications from source allows fine-tuning
of build output to the hardware to make maximum use of energy saving
optimizations.

Backing up the system by creating a full bit-for-bit copy is a
time-consuming process which often leads to infrequent backups and, as a
result, occasional data loss due to storage card failure. SkiffOS
provides an alternate approach, in which existing OS distributions can
be imported as containers. Containers are easy to back up, restore,
roll-back, and are portable between machines of similar architecture.
Any vendor-provided BSP can be imported as a container and used without
sacrifices to workflow or compatibility.

\hypertarget{nixos}{%
\subsubsection{NixOS}\label{nixos}}

NixOS \autocite{nixos} argues that imperative package managers
destructively update the state of the system, leading to "inability to
roll back changes easily, to run multiple versions of a package
side-by-side, to reproduce a configuration deterministically on another
machine, or to reliably upgrade a system." NixOS includes a declarative
package manager which performs modifications according to a
specification of target system state. Nix can perform atomic upgrades
with the ability to roll-back.

SkiffOS is designed around the same principles of immutability and
declarative configuration, but provides these guarantees through
compilation of the system in advance, loading an immutable and ephemeral
root system at boot-up. It also addresses the other issues described by
the NixOS paper, including running multiple systems side-by-side in
Linux containers, and reproducing a system configuration
deterministically at a later time.

NixOS breaks the Filesystem Hierarchy Standard (FHS) \autocite{fhs}
declared by the Linux foundation and followed by most distributions,
instead using a flat symbolic link-based structure. This causes
incompatibility with software not compiled by Nix. On the other hand,
Buildroot \autocite{buildroot} (and therefore SkiffOS) follow the FHS
for compatibility with existing glibc-based binaries.

SkiffOS focuses on providing a minimal "shim" to abstract away the
differences between hardware (particularly Single-Board Computers) such
that the containers are portable to new platforms. It does not
distribute complex packages like web browsers, graphical interfaces, and
other user applications. This responsibility is left to the
distributions running in Linux containers attached to persistent storage
and/or customization of the cross-compiled system.

NixOS has been integrated with SkiffOS as a "core" container
configuration. This approach uses Buildroot to manage the hardware
specifics, and NixOS to handle declarative system configuration and
rolling application upgrades in containers. Software not compiled by Nix
can still be run in a concurrent container.

\hypertarget{buildroot}{%
\subsubsection{Buildroot}\label{buildroot}}

The Buildroot \autocite{buildroot} project provides a automated system
cross-compiler. Users typically download a Buildroot release, create a
configuration by hand, and store this along with their other project
files. It focuses on providing a toolset for Embedded Linux developers
to produce system images for use with a single target platform and/or
product.

SkiffOS extends Buildroot to simplify this process with an easy to
understand configuration layering architecture, which merges together
the platform support configs with the selected components to configure
the build automatically. The configuration layers can be stored
externally to the SkiffOS tree. Existing Buildroot projects can be
adapted as configuration layers and ported to any of the available
target configurations.

\hypertarget{yocto-project}{%
\subsubsection{Yocto Project}\label{yocto-project}}

The Yocto Project \autocite{yocto} is closely comparable to Buildroot,
and is primarily focused on the python-based "bitbake" tool, which is
derived from Gentoo's "portage," with a focus on cross-compilation of
complex embedded Linux systems. This is contrasted with Buildroot's
focus on simplicity, using the Makefile and Kconfig architecture
\autocite{yoctobr}. Yocto system configurations are fragmented into
"overlay" repositories from various sources, while Buildroot focuses on
a consolidated and curated package tree with strict and opinionated code
style. Yocto has many more packages than Buildroot.

Buildroot \autocite{buildroot} forms a linear commit history with
release checkpoints, and Skiff releases pin the version of the Buildroot
sub-module along with the versions of the device firmware, kernels, and
board support files. A given Git checkout is reproducible forever
\autocite{buildrootrt}. Yocto setups may suffer from loss of
availability of overlay repositories, or de-synchronization between
changes to the overlay repositories and changes to the core system
configurations.

\hypertarget{conclusion}{%
\section{Conclusion}\label{conclusion}}

SkiffOS offers reproducible system behavior, offline compilation,
immutable host system for containers attached to persistent storage,
board-specific build re-targeting via config layers, and Over-The-Air
upgrade/downgrade. The host system and/or containers can be easily
customized for specific use cases with cross-platform configuration
layers. The "Skiff Core" workflow provides an easy way to import any
existing Linux userspace such that users will likely not realize they
are now working in a container. Background workloads can run in parallel
Linux containers, independently isolated from the user workspace(s)
including resource management and quality-of-service (QoS).

SkiffOS is available under the MIT license at:

\url{https://github.com/skiffos/skiffos}

Buildroot is available under the GPLv2 license at:

\url{https://buildroot.org}

\clearpage

\printbibliography

\end{document}